\documentclass[11pt,a4paper]{article}
\usepackage{jheppub}
\usepackage{multirow}
\usepackage{braket}
\usepackage{bbm}

\title{Limits on $CP$-violating hadronic interactions and proton EDM from paramagnetic molecules}

\author[a,b]{V. V. Flambaum}
\author[a]{I. B. Samsonov}
\author[a]{H. B. Tran Tan}

\affiliation[a]{School of Physics, University of New South Wales,
Sydney 2052, Australia}
\affiliation[b]{Helmholtz Institute Mainz, Johannes Gutenberg University, 55099 Mainz, Germany}

\emailAdd{v.flambaum@unsw.edu.au}
\emailAdd{igor.samsonov@unsw.edu.au}
\emailAdd{tan.tran@student.unsw.edu.au}

\abstract{
Experiments with paramagnetic ground or metastable excited states of molecules (ThO, HfF$^+$, YbF, YbOH, BaF, PbO, etc.) provide strong constraints on electron electric dipole moment (EDM) and coupling constant $C_{SP}$ of contact semileptonic interaction.  We compute new contributions to $C_{SP}$ arising from the nucleon EDMs due to combined electric and magnetic electron-nucleon interaction. This allows us to improve limits from the experiments with paramagnetic molecules on the $CP$-violating parameters, such as the proton EDM, $|d_p|< 1.1\times 10^{-23} e\cdot $cm, the QCD vacuum angle, $|\bar \theta|<1.4\times 10^{-8}$, as well as the quark chromo-EDMs and $\pi$-meson-nucleon couplings. Our results may also be used to search for the axion dark matter which produces oscillating $\bar\theta$.}

\makeatletter
\gdef\@fpheader{\phantom{a}}
\makeatother

\arxivnumber{2004.10359}

\begin{document}

\maketitle

\section{Introduction} The Standard Model of elementary particles naturally incorporates the sources for $CP$ (charge and parity) violation represented by the Cabibbo–Kobayashi–Maskawa (CKM) matrix \cite{C,KM} and the QCD vacuum angle\ \cite{Peccei1977,Weinberg1978,Wilczek1978} (see also Refs.\ \cite{yamanaka2017,chupp2019} and further references therein). While the elements of the CKM matrix are measured to a high accuracy, the exact value of the QCD vacuum angle $\bar{\theta}$ is not known. In recent years, the precision in modern atomic and molecular EDM experiments has been improved to such a level that constraints on $\bar{\theta}$ and other $CP$-violating parameters imposed by these experiments are approaching or even exceeding those of particle physics\ \cite{Griffith2009,Baker2006,Hudson2011,Baron2014,Parker2015,Pendlebury2015,Graner2016}. For example, experiments with diamagnetic atoms and molecules targeting the nuclear Schiff moments have placed significant bounds on the nucleon EDMs 
\cite{Harrison1969,Hinds1980,Wilkening1984,Schropp1987,Cho1991,Vold1984,Chupp1994,Stoner1996,Bear1998,
Rosenberry2001,romalis2001,Griffith2009,Graner2016,Parker2015,Bishof2016} whereas those using paramagnetic polar molecules \cite{Hudson2011,Loh2013,PbO2013,HfF2017,BaF2018,ACMEII} give rise to the most stringent constraints on the electron EDM.

In paramagnetic atoms, an atomic EDM may be induced by the following contact $CP$-odd semileptonic operators
\begin{equation}\label{contact-interaction}
{\cal L} = \frac{G_F}{\sqrt2}  C_{SP}^p\, \bar e i \gamma_5 e\,
\bar p p 
+ \frac{G_F}{\sqrt2} C_{SP}^n\, \bar e i \gamma_5 e\,
\bar n n \,,
\end{equation}
where $G_F$ is the Fermi coupling constant, $e$, $p$ and $n$ are respectively the electron, proton and neutron fields; $C_{SP}^p$ and $C_{SP}^n$ are the electron couplings to the proton and neutrons, respectively. The subscript $SP$ denotes the nucleon-scalar and electron pseudoscalar two-fermion bilinears. 

In polarised polar molecules, the interaction\ \eqref{contact-interaction} induces shifts of energy levels. The measurement of these shifts places constraints on the value of $C_{SP}\equiv C_{SP}^p Z/A + C_{SP}^n N/A$, where $A$ and $Z$ are the nuclear mass and charge numbers, and $N=A-Z$ is the number of neutrons. The most stringent constraint on $C_{SP}$ is placed by the ACME collaboration\ \cite{ACMEII}, which used the molecule $^{232}{\rm ThO}$, (90\% C.L.)
\begin{equation}
    \left|C_{SP}\right|_{\rm Th}
< 7.3 \times 10^{-10}\,.
\label{Csp-constr}
\end{equation}

The coupling constant $C_{SP}$ receives contributions from various sources, which include interactions with the nucleon EDM $d_{p,n}$. The parameters $d_{p,n}$ may, in turn, be expressed in terms of more fundamental ones, namely, the $CP$-odd $\pi$-meson-nucleon coupling constants $\bar g^{(0,1)}_{\pi NN}$, the quark chromo-EDMs $\tilde{d}_{d,u}$ and the QCD vacuum angle $\bar{\theta}$. Our aim is to determine the leading dependence of $C_{SP}$ on the parameters $d_{p,n}$, $\bar g^{(0,1)}_{\pi NN}$, $\tilde{d}_{d,u}$ and $\bar{\theta}$ for ${}^{232}$Th and several other atoms of experimental interest, including Ba, Yb, Hf, Pb and Ra. Note that $CP$-violating effects rapidly increase with nuclear charge (see next sections). Therefore, in molecules the effects come from the heaviest nucleus.

In a recent paper \cite{FPRS}, the contributions to $C_{SP}$ from the two-photon and $\pi,\eta$-meson exchanges between electrons and nucleons were calculated. These contributions led to a limit on the QCD vacuum angle $|\bar{\theta}|\lesssim 3\times 10^{-8}$. In this paper, we take into account additional contributions to  $C_{SP}$ which are comparable or even bigger than those calculated in Ref.~\cite{FPRS}, namely, contribution of nucleon transitions in discrete spectrum (which are enhanced by small energy denominators in perturbation theory). We take into account the effect of the  Coulomb interaction in intermediate electron states. In heavy atoms like thorium with $Z=90$, the parameter $Z \alpha \sim 1$ and the Coulomb interaction is important.
 
As we will demonstrate further, the addition of the new contributions allows us to significantly improve the limits on the QCD vacuum angle and other $CP$-violating hadronic parameters. We stress that these results play important role in the search of axion-like dark matter \cite{Roussy2020} which manifests itself as  oscillating $\bar\theta$-term (see, e.g., Ref.\ \cite{Graham-Rajendran}).

It is important to note that, according to a theorem by Schiff \cite{Schiff}, atomic electrons completely shield the nuclear EDM. However, it can be measured through the nuclear Schiff moments and magnetic quadrupole moments \cite{Schiffmoment1,Schiffmoment2,quadrupole}, or by means of applying an oscillating electric field and observing nuclear spin rotations as argued in the Refs.~\cite{Dzuba86,Victor18,my1,my2,my3}. However, since the Schiff theorem applies only to a system which interacts electrically, the interaction of the atomic electrons with the magnetic dipole moment of the nucleus allows for a non-zero atomic EDM induced by a nuclear EDM \cite{Schiff}. This paper is devoted to the study of the mechanism for the production of the atomic EDM from the combined electric and magnetic interactions between the atomic electrons and the nucleus.

In Ref.~\cite{Ginges}, it was argued that for atoms with vanishing nuclear spins there are no non-vanishing contributions to the atomic EDM from $CP$-violating nuclear scalar polarizability. However, the analysis of Ref.\ \cite{Ginges} did not take into account specific near-nucleus electronic contributions which cannot be reduced to the $CP$-violating nuclear scalar polarizability and are significantly enhanced by relativistic effects in heavy atoms. Indeed, since the electronic $s$ and $p$ Dirac wave functions for a point-like nucleus are singular at the origin, the electronic matrix elements between these states are formally divergent and thus make significant contributions to the induced atomic EDM. In this paper, we will systematically analyze the contributions to the atomic EDM from such matrix elements in atoms with vanishing nuclear spins and compare them with those arising from the contact electron-nucleon interaction \eqref{contact-interaction}. This will allow us to obtain the leading-order dependence $C_{SP}=C_{SP}(d_{p,n},\bar g^{(0,1)}_{\pi NN},\tilde{d}_{d,u},\bar{\theta})$ for several atoms of experimental interest and deduce improved limits on the $CP$-violating hadronic parameters.

The rest of the paper is organized as follows. In Sect.~\ref{Atomic EDM due to contact electron-nucleon interaction}, we present an estimate for the atomic EDM arising from the contact electron-nucleon interaction. Section \ref{Contribution to the atomic EDM from nucleon permanent EDMs} is devoted to the computation of the atomic EDM induced by nucleon permanent EDMs. In Sect.~\ref{Constraints}, we compare the contributions to the atomic EDM from the contact electron-nucleon interaction and the nucleon permanent EDMs and find relations between the constant $C_{SP}$ and $CP$-violating hadronic parameters. In Sect.~\ref{Conclusion} we give a summary of our results and provide some comments on assumptions and precision. Technical details of calculations of electronic and nuclear matrix elements are collected in appendices.

Throughout this paper we use natural units with $c=\hbar=1$.

%%%%%%%%%%%%%%%%%%%%%%%%%%%%%%%%%%%%%%%%%%%%%%%%%%%%%%%%%%%%%%%%%%%%%%%%%%%%%%%%%%%%%%

\section{Atomic EDM due to contact electron-nucleon interaction}
\label{Atomic EDM due to contact electron-nucleon interaction} 
In an atom, the $CP$-odd interaction\ \eqref{contact-interaction} between a valence electron and the nucleus is described by the Hamiltonian \cite{NewBounds1985}
\begin{equation}
H_{\rm cont}=\frac{iG_F}{\sqrt{2}}AC_{SP}\gamma_0\gamma_5 \rho({\bf R})\,,
\label{Hcont}
\end{equation}
where $\gamma_0=\left( \begin{smallmatrix} 
\mathbbm{1}_{2\times2} & 0 \\ 0 & -\mathbbm{1}_{2\times2}
\end{smallmatrix}\right)$ 
and $\gamma_5=
\left( \begin{smallmatrix} 
0& -\mathbbm{1}_{2\times2}  \\  -\mathbbm{1}_{2\times2} &0
\end{smallmatrix}\right)$ are the Dirac matrices, $\bf R$ is the position vector of the electron and $\rho({\bf R})$ is the normalized nuclear charge density. In the leading approximation, $\rho({\bf R})$ is constant inside the nucleus of radius $R_0$ and vanishes outside, $\rho({\bf R})=3\theta(R_0-R)/(4\pi R_0^3)$, where $\theta(x)$ is the Heaviside step function.

Matrix elements of the operator (\ref{Hcont}) receive non-vanishing contributions only from small distances, where the nuclear density $\rho({\bf R})$ is different from zero. In heavy nuclei, the $s_{1/2}$ and $p_{1/2}$ electron wave functions have large relativistic enhancement inside the nucleus, as compared with other wave functions corresponding to higher angular momenta $l$ which are negligible at the nucleus ($\propto R^l$) \cite{NewBounds1985}. As a result, the atomic EDM receives dominant contributions from the matrix element of the operator \eqref{Hcont} with the $s_{1/2}$ and $p_{1/2}$ states, 
\begin{equation}
\label{de_contact}
{\bf d}\approx 2\frac{\bra{s_{1/2}}e{\bf R}\ket{p_{1/2}}\bra{p_{1/2}}H_{\rm cont}\ket{s_{1/2}}}{E_{p_{1/2}}-E_{s_{1/2}}}\,,
\end{equation}   
where $E_{{s}_{1/2}}$ and $E_{{p}_{1/2}}$ are the energies of the $s_{1/2}$ and $p_{1/2}$ states, respectively, and $-e$ is the electron charge.  

Here, we will estimate the matrix element $\bra{p_{1/2}}H_{\rm cont}\ket{s_{1/2}}$ for heavy atoms. In subsequent sections, this matrix element will be compared with those of $CP$-odd operators originating from nucleon EDMs.
   
For an atom with a point-like nucleus, the $s_{1/2}$ and $p_{1/2}$ valence electron wave functions have simple analytic expressions, see Eqs.\ \eqref{Bessel}. For an extended nucleus model with constant charge density, it is sufficient to consider a simple continuation of the wave functions \eqref{Bessel} to the region inside the nucleus as
\begin{subequations}\label{psi-ass}
\begin{eqnarray}
\ket{s_{1/2}} &= &c_{s_{1/2}} \frac{ R_0^{\gamma-1}}{\Gamma(2\gamma+1)} \left(\frac{2Z}{a_B}\right)^{\gamma}\left(
\begin{array}{c}
-(\gamma+1) \Omega^{-1}_{\mu } \\
\frac{iZ\alpha R}{R_0}\Omega^1_{\mu}
\end{array}
\right)\,,\\
\ket{p_{1/2}} &= & c_{p_{1/2}}\frac{ R_0^{\gamma-1}}{\Gamma(2\gamma+1)} \left(\frac{2Z}{a_B}\right)^{\gamma} \left(
\begin{array}{c}
\frac{(1-\gamma)R}{R_0} \Omega^1_{\mu } \\
iZ\alpha \Omega^{-1}_{\mu}
\end{array}
\right)\,,
\end{eqnarray}
\end{subequations}
where $\Omega^{\kappa }_{\mu}$ is the spherical spinor, $a_B$ is the Bohr radius, $\gamma=\sqrt{1-Z^2\alpha^2}$ is the relativistic factor and $\alpha\approx1/137$ is the fine structure constant. The values of the normalization constants $c_{s_{1/2}}$, $c_{p_{1/2}}$ are not specified here, as our final results will be independent from these constants. 

With the wave functions \eqref{psi-ass}, the matrix element of the operator \eqref{Hcont} reads
\begin{equation}   
\bra{p_{1/2}}H_{\rm cont}\ket{s_{1/2}} = -c_{s_{1/2}}c_{p_{1/2}}\frac{G_F C_{SP}}{10\sqrt2\pi}\frac{1+4\gamma}{\Gamma(2\gamma+1)^2}\frac{AZ\alpha}{R_0^2} 
\left(\frac{2ZR_0}{a_B}\right)^{2\gamma}.
\label{matrix-contact}
\end{equation}
For heavy nuclei, this matrix element is significantly enhanced due to the factor $AZ^{2\gamma+1}$. 
Since we will compare Eq.\ \eqref{matrix-contact} with the contribution from nucleon EDMs, which is not enhanced that strongly, it appears that for the ratio of the effects (which is the contribution from nucleon EDMs to the effective constant $C_{SP}$) lighter nuclei may have bigger $C_{SP}$ than heavy ones.

%%%%%%%%%%%%%%%%%%%%%%%%%%%%%%%%%%%%%%%%%%%%%%%%%%%%%%%%%%%%%%%%%%%%%%%%%%

\section{Contribution to the atomic EDM from nucleon permanent EDMs}\label{Contribution to the atomic EDM from nucleon permanent EDMs}

In this section, we calculate the contribution to the atomic EDM originating from permanent nucleon EDMs. Since electric interactions alone cannot give rise to atomic EDMs due to the Schiff theorem \cite{Schiff}, such a contribution can only arise when both the electric and magnetic electron-nucleon interactions are taken into account. We start with the review of the effective Hamiltonian of this interaction and in subsequent subsections calculate the nuclear and electronic matrix elements of this Hamiltonian.

\subsection{Effective Hamiltonian for the \texorpdfstring{$CP$}{}-odd electron-nucleon interaction}
Let ${\bf d}_i = d_i \boldsymbol{\sigma}_i$ and $\boldsymbol{\mu}_i=\mu_0 (g_i^l {\bf l}_i + g_i^s {\bf s}_i)$ be the operators of electric and magnetic dipole moments of the $i$-th nucleon in the nucleus. Here $d_i = d_{p,n}$ are the proton and neutron permanent EDMs, $\mu_0$ is the nuclear magneton, $g_{i}^l=g_{p,n}^l$ and $g_{i}^s=g_{p,n}^s$ are the orbital and spin $g$-factors of the nucleons. The operators ${\bf d}_i$ and $\boldsymbol{\mu}_i$ couple with the electric and magnetic fields of the valence electron, yielding the interaction Hamiltonian\footnote{In general, in addition to the nucleon EDMs, there are other $CP$-violating operators at hadronic level including three-pion and four-nucleon couplings. In Ref.~\cite{Bsaisou}, contributions to the nuclear EDM due to these operators were found for some light nuclei. In this paper, we focus only on the contributions to the atomic EDM due to the nucleon permanent EDMs.}
\begin{subequations}\label{mix_H}
\begin{align}
H&=-\sum_{i=1}^A\left(H^d_i+H^\mu_i\right)\,,\label{HInt}\\
H^d_i&=\frac{e\,{\bf d}_i\cdot\left({\bf R}-{\bf r}_{i}\right)}{\left|{\bf R}-{\bf r}_{i}\right|^3}\,,\label{HE}\\
H^\mu_i&=\frac{e\,\boldsymbol{\mu}_i\cdot\left[\left({\bf R}-{\bf r}_{i}\right)\times\boldsymbol{\alpha}\right]}{\left|{\bf R}-{\bf r}_{i}\right|^3}\,,
\label{HB}
\end{align}
\end{subequations}
where ${\bf r}_{i}$ are the position vectors of the nucleons and $\boldsymbol{\alpha}=
\left(\begin{smallmatrix}
0 & \boldsymbol{\sigma} \\ \boldsymbol{\sigma} & 0
\end{smallmatrix}\right)$ are Dirac matrices acting on electron wave functions. 

The unperturbed atomic states will be denoted by $|mm'\rangle= |m\rangle|m'\rangle$, where $|m\rangle$ and $|m'\rangle$ are electronic and nuclear states, respectively.\footnote{In what follows, the nuclear quantum number will be distinguished from the electronic ones with the apostrophe.} The first-order contributions to the atomic EDM due to the interaction Hamiltonian\ \eqref{mix_H} vanish for spinless nuclei which we consider in this paper, $\langle 0' | {\bf s}|0'\rangle=0$. The leading non-vanishing contributions to the atomic EDM arise in the second order of the perturbation theory (see Ref.\ \cite{Ginges} for further details), 
\begin{equation}
{\bf d}= -2\sum_{m,n,n'}\frac{\bra{0}e\mathbf{R}\ket{m}\bra{0'm}H\ket{nn'}\bra{n'n}H\ket{00'}}{\left(E_{m}-E_{0}\right)\left[\Delta E_{n}+{\rm sgn}(E_n)\Delta E_{n'}\right]}\,,\label{datom}
\end{equation}
where the sum is taken over the excited states with $m\ne 0$ and $nn'\ne 00'$. Here, $E_n$ and $E_{n'}$ are energies of electronic and nuclear excitations, respectively, whereas $\Delta E_{n}\equiv E_{n}-E_{0}$ and $\Delta E_{n'}\equiv E_{n'}-E_{0'}$. Note that the sum over the intermediate electronic states $\ket{n}$ includes both positive and negative energy levels which are inherent in Dirac's theory. The negative energy states contribute to Eq.~(\ref{datom}) with opposite sign of $\Delta E_{n'}$ because they may be interpreted as blocking contributions for the electrons from the Dirac sea which cannot be excited to the occupied electron orbitals.  Therefore, contribution of the transitions from the Dirac sea must be subtracted. (See also Refs.~\cite{Plunien91,Pachucki93,Plunien95} for analogous account of the negative energy states within the problem of atomic energy shift due to nuclear polarizability.)

As is argued in Sect.~\ref{Atomic EDM due to contact electron-nucleon interaction}, the atomic EDM \eqref{datom} receives leading contributions from the matrix elements with $\ket{0}=\ket{s_{1/2}}$ and $\ket{m}=\ket{p_{1/2}}$. As a result, Eq.\ \eqref{datom} may be cast in the form  \eqref{de_contact}, with the contact interaction operator $H_{\rm cont}$ replaced with an effective interaction Hamiltonian $H_{\rm eff}$ defined by 
\begin{equation}\label{EffectiveMatrixElement}
H_{\rm eff}\equiv -\sum_{nn'\ne 00'}\frac{\ket{m}\bra{0'm}H\ket{nn'}\bra{n'n}H\ket{00'}\bra{0}}{\Delta E_{n}+{\rm sgn}(E_n)\Delta E_{n'}}\,.
\end{equation}
Substituting the operators \eqref{mix_H} into Eq.~\eqref{EffectiveMatrixElement}
and keeping only linear in ${\bf d}_i$ terms we find
\begin{equation}\label{Scalar contribution}
\bra{p_{1/2}}H_{\rm eff}\ket{s_{1/2}}=-\sum_{n' \ne 0'}\sum_{i,j=1}^A\mathcal{M}^{n'}_{ij}\,,
\end{equation}
where $\mathcal{M}^{n'}_{ij}$ is
\begin{equation}\label{ME}
\begin{aligned}
\mathcal{M}^{n'}_{ij}
&=\sum_{n}\frac{\bra{p_{1/2}0'}H^\mu_i\ket{n'n}\bra{nn'}H^d_j\ket{0's_{1/2}}}{\Delta E_{n}+{\rm sgn}(E_n)\Delta E_{n'}}+(s_{1/2}\leftrightarrow p_{1/2})\,.\\
\end{aligned}
\end{equation}
In subsequent subsections, we will compute the nuclear and electronic matrix elements in this expression and present the results for the matrix element (\ref{Scalar contribution}).

\subsection{Integration over radial nuclear coordinates}\label{IIIB}
The form of the nuclear matrix elements $\bra{0'}H^\mu_i\ket{n'}$ and $\bra{0'}H^d_i\ket{n'}$ depends on the specific nuclear model. To compute these matrix elements we consider an approximation in which the nuclear wave functions are given by
\begin{equation}\label{NucWave}
\ket{n'}=\prod_{i=1}^A R_{n'}(r_i)\Theta_{n'}(\hat{\bf r}_i)\,,
\end{equation}
where $\hat{\bf r}_i={\bf r}_i/r_i$ is the unit vector pointing in the direction of ${\bf r}_i$ and $R_{n'}(r_i)$ is the radial wave function.

The function $\Theta_{n'}(\hat{\bf r}_i)$ in Eq.~(\ref{NucWave}) specifies the angular dependence of $\ket{n'}$. For a spherical nucleus, it is convenient to use the spherical basis, in which the nucleon states have definite total angular momenta $j_i^{n'}$ and total magnetic quantum numbers $\mu_i^{n'}$. In this case, one may take $\Theta_{n'}(\hat{\bf r}_i)=\Omega^{\kappa_i^{n'}}_{\mu_i^{n'}}(\hat{\bf r}_i)$ where $\Omega^{\kappa_i^{n'}}_{\mu_i^{n'}}$ is the two-component spherical spinor. On the other hand, for deformed nuclei, one usually employs the Nilsson basis \cite{Nilsson}, in which nucleon states have  
orbital magnetic quantum numbers $m_i^{n'}$ and spin projections $s_i^{n'}=\pm 1/2$. 

In the spin-flip matrix elements produced by the nucleon magnetic moment and EDM operators, the radial wave functions in the bra and ket states are the same and we may average over their oscillations, taking\footnote{We stress that the function 
(\ref{R}) is needed only to generalize the point-like interaction operators (\ref{mix_H}) to the model with extended nucleus. Upon averaging the operators (\ref{mix_H}) over nuclear density (\ref{R}) the short-distance singularity of the electronic operators is spread and effectively regularized. Physically this means that the electron interacts with a nuclear charge distribution which has no singularity at the origin. One could model the nuclear charge distribution with the more realistic Woods-Saxon formula. However, we have checked that the use of the Woods-Saxon formula changes the electronic matrix elements by about 1\% which is not essential within this work.}
 \begin{equation}
\label{R}
    |R_{n'}(r_i)|^2 =\frac{3}{R_0^3}\theta(R_0-r_{i})\,. 
\end{equation}
Using the radial function (\ref{R}), one may integrate over the radial variable $ r_i$ in the matrix elements $\bra{0'}H^\mu_i\ket{n'}$ and $\bra{0'}H^d_i\ket{n'}$, keeping only the leading dipole terms with respect to the electronic coordinate $\bf R$. For that purpose, we note that the operators (\ref{HE}) and (\ref{HB}) depend on the function
\begin{equation}\label{Nabla}
\frac{{\bf R}-{\bf r}_i}{|{\bf R}-{\bf r}_i|^3}=-\nabla_{\bf R}\frac1{|{\bf R}-{\bf r}_i|}
=-\nabla_{\bf R}\sum_{L=0}^\infty \frac{r^L_< }{ r^{L+1}_>}P_L(\cos\varphi)\,,
\end{equation}  
where $r_<\equiv\min(R,r_i)$, $r_>\equiv\max(R,r_i)$ and $\varphi$ is the angle between $\bf R$ and ${\bf r}_i$. In this expansion, it is sufficient to keep only the $L=0$ term which reads $\theta(R-r_i)/R+\theta(r_i-R)/r_i$, because after applying the derivative $\nabla_{\bf R}$ it amounts to the dipole term with respect to the electronic coordinate $\bf R$. Taking into account only this term and integrating over the nuclear coordinates $r_i$, one may cast the nuclear matrix elements $\bra{0'}H^\mu_i\ket{n'}$ and $\bra{0'}H^d_i\ket{n'}$ in the form
\begin{subequations}\label{16}
\begin{eqnarray}
\bra{0'}H^\mu_i\ket{n'}&=&e\textbf{M}\cdot\bra{\Theta_{0'}}\boldsymbol{\mu}_i\ket{\Theta_{n'}}\,,\label{Fmu1}\\
\bra{0'}H^d_i\ket{n'}&=&e\textbf{D}\cdot\bra{\Theta_{0'}}{\bf d}_i\ket{\Theta_{n'}}\,,\label{Fd1}
\end{eqnarray}
\end{subequations}
where the operators $\textbf{M}$ and $\textbf{D}$ act only on electronic variables 
\begin{subequations}\label{EffectiveOps}
\begin{eqnarray}
\textbf{M}&\equiv&
\left[\theta(R-R_0)\frac{1}{R^2}+\theta(R_0-R)\frac{R}{R_0^3} \right](\hat{\bf R}\times\boldsymbol{\alpha})\,,\label{Fmu2}\\
\textbf{D}&\equiv&
\left[ \theta(R-R_0)\frac{1}{R^2}+\theta(R_0-R)\frac{R}{R_0^3}\right]\hat{\bf R}\,.\label{Fd2}
\end{eqnarray}
\end{subequations}
Here $\hat{\bf R}={\bf R}/R$ is the unit vector pointing in the direction of $\bf R$.

Substituting Eqs.\ \eqref{16} into Eq.\ \eqref{ME}, one may factorize the matrix element $\mathcal{M}^{n'}_{ij}$ into nuclear and electronic parts as
\begin{subequations}
\label{3.12}
\begin{align}
\mathcal{M}^{n'}_{ij}&=c_{s_{1/2}}c_{p_{1/2}}d_i\,\Xi^{n'}_{i}M(\Delta E_{n'})\delta_{ij}\,,\label{Diagonal}\\
\Xi^{n'}_{i}&=\bra{\Theta_{0'}}\boldsymbol{\mu}_i\ket{\Theta_{n'}}\bra{\Theta_{n'}}\boldsymbol{\sigma}_i\ket{\Theta_{0'}}\,,\label{beta-n}\\
M(\Delta E_{n'})
&=\frac{\alpha}{3c_{s_{1/2}}c_{p_{1/2}}}\sum_{n}\frac{\bra{p_{1/2}}{\bf M}\ket{n}\bra{n}\textbf{D}\ket{s_{1/2}}}{\Delta E_{n}+{\rm sgn}(E_n)\Delta E_{n'}}+(s_{1/2}\leftrightarrow p_{1/2})\,,\label{EMatrix}
\end{align}
\end{subequations}
where we have separated the normalization constants $c_{s_{1/2}}$ and $c_{p_{1/2}}$ out of the electronic matrix element $M(\Delta E_{n'})$. 

Equations \eqref{3.12} deserve some comments. First, we note that the matrix element \eqref{Diagonal} is represented by a diagonal matrix with respect to nucleon indices $ij$. Indeed, the two matrix elements in the right-hand side\ of Eq.~(\ref{beta-n}) should involve the same nucleon because they both have the same initial and final nuclear states $|0'\rangle$ and $|n'\rangle$. Secondly, in passing from Eqs.~(\ref{16}) to Eqs.~(\ref{3.12}), we have implemented the substitution $({\bf M}\cdot \boldsymbol{\mu})({\bf D}\cdot {\bf d}) \to \frac13 ({\bf M}\cdot {\bf D})(\boldsymbol{\mu} \cdot  {\bf d})$ which is applicable inside the matrix elements. Note that the nuclear matrix element (\ref{beta-n}) is scalar for spinless nuclei which we consider in this paper.

For each nuclear energy $\Delta E_{n'}$, the value of the function $M(\Delta E_{n'})$ in Eq.~(\ref{EMatrix}) is calculated by directly summing over the intermediate electronic states $|n\rangle$, which occupy both the discrete and continuum spectra. The contribution from the discrete spectrum appears negligible as compared with the continuum one (see Refs.~\cite{Plunien91,Plunien95} for analogous calculations). Therefore, in what follows, we will consider intermediate electronic states $|n\rangle$ from the continuum spectrum represented by Dirac-Coulomb wave functions described in Appendix \ref{AppB2}.

\subsection{Nuclear spin-flip matrix elements for spherical and deformed nuclei}\label{Nuclear spin-flip}
The nuclear matrix elements \eqref{beta-n} correspond to M1 spin-flip nuclear transitions. The computations of these matrix elements slightly differ for spherical and deformed nuclei. For the former, one uses $\Theta_{n'}(\hat{\bf r}_i)=\Omega^{\kappa_i^{n'}}_{\mu_i^{n'}}(\hat{\bf r}_i)$ and the matrix elements of the total momentum ${\bf j}_i$ vanish for fine structure doublets. As a result, substituting the operators ${\bf d}_i$ and $\boldsymbol{\mu}_i$ into Eq.~\eqref{beta-n} and using the identity ${\bf l}={\bf j}-{\bf s}$, the nuclear matrix elements may be written as $\Xi^{n'}_{i}= 2\mu_0(g^s_i - g^l_i)\left|\bra{\Theta_{n'}}{\bf s}_{i}\ket{\Theta_{0'}}\right|^2$. For deformed nuclei, the angular functions may be chosen as $\Theta_{n'}(\hat{\bf r}_i)=Y^{l_i^{n'}}_{m_i^{n'}}(\hat{\bf r})\zeta_{\pm1/2}$, where $Y^{l_i^{n'}}_{m_i^{n'}}(\hat{\bf r})$ are spherical harmonics and $\zeta_{\pm1/2}$ is a two-component unit spinor. In this case, one may verify that $\bra{\Theta_{0'}}{\bf l}_i\ket{\Theta_{n'}}\bra{\Theta_{n'}}{\bf s}_i\ket{\Theta_{0'}}=0$. Thus, for deformed nuclei, in Eq.\ \eqref{beta-n} we have $\Xi^{n'}_{i}= 2\mu_0g^s_i\left|\bra{\Theta_{n'}}{\bf s}_i\ket{\Theta_{0'}}\right|^2$. These two cases may be combined in one expression
\begin{equation}
\Xi^{n'}_{i}=2 \mu_0 \left(g_{i}^s- \epsilon g_{i}^l\right)|\bra{\Theta_{0'}}{\bf s}_i\ket{\Theta_{n'}}|^2\,,
\label{beta-n-spherical}
\end{equation}
where 
\begin{equation}
    \epsilon = \left\{
    \begin{array}{ll}
        1\quad&\mbox{for spherical nuclei} \\
        0 & \mbox{for deformed nuclei.}
    \end{array}
    \right.
\end{equation}

In Eq.\ \eqref{beta-n-spherical}, the values of the $g$-factors for the proton and neutron are $g_p^l = 1$, $g_p^s = 5.586$, $g_n^l = 0$ and $g_n^s = -3.826$, respectively. The energies $\Delta E_{n'}$ and matrix elements $\bra{\Theta_{0'}}{\bf s}_i\ket{\Theta_{n'}}$ of the M1 spin-flip transitions are listed in Appendix \ref{NuclearSection} for different nuclei of interest. 

\subsection{Matrix element of the effective Hamiltonian}
It is convenient to denote the energies and matrix elements for proton M1 spin-flip transitions as $\Delta E_p$ and $\bra{0'}{\bf s}\ket{n'}_p$, and for neutron transitions as $\Delta E_n$ and $\bra{0'}{\bf s}\ket{n'}_n$. Substituting the nuclear matrix elements \eqref{beta-n-spherical} into Eq.~(\ref{Diagonal}), the matrix element \eqref{Scalar contribution} may be presented in the form
\begin{equation}
    \bra{p_{1/2}} H_{\rm eff}\ket{s_{1/2}}
    =-2c_{s_{1/2}}c_{p_{1/2}}\mu_0\left[ d_p (g_p^s-\epsilon g_p^l) M_p
    + d_n (g_n^s-\epsilon g_n^l ) M_n\right]\,,
\label{matrix-d}
\end{equation}
where the quantities $M_p$ and $M_n$ are defined by
\begin{subequations}\label{MpMnDef}
\begin{eqnarray}
M_{p} &\equiv&  \sum_{\Delta E_p}|\bra{0'}{\bf s}\ket{n'}_p|^2 M(\Delta E_p)\,,\\
M_{n} &\equiv&  \sum_{\Delta E_n}|\bra{0'}{\bf s}\ket{n'}_n|^2 M(\Delta E_n)\,.
\label{Mpn}
\end{eqnarray}
\end{subequations}
Note that for deformed nuclei Eqs.~(\ref{matrix-d}) and (\ref{MpMnDef}) simplify, because in the Nilsson basis $|\bra{0'}{\bf s}\ket{n'}_{p,n}|^2=1$ and $\epsilon=0$.

The details of the numerical computations of the coefficients $M_p$ and $M_n$ are given in Appendix \ref{AppB} and the results are presented in Table~\ref{Electronic}. In particular, for $^{180}{\rm Hf}$ and $^{232}$Th they are: 
\begin{equation}\label{3.16}
\begin{aligned}
 ^{180}{\rm Hf}: &\quad &     M_{p}=121a_B^{-1}\,,\qquad M_{n}=114a_B^{-1}\,,\\
 ^{232}{\rm Th}: &\quad &     M_{p}=244a_B^{-1}\,,\qquad M_{n}=385a_B^{-1}\,.
\end{aligned}
\end{equation}  
Substituting these values into Eq.~(\ref{matrix-d}), we find
\begin{subequations}\label{317}
\begin{align}
\bra{p_{1/2}} H_{\rm eff}\ket{s_{1/2}}_{{}^{180}{\rm Hf}} &= -2 c_{s_{1/2}}c_{p_{1/2}}\frac{\mu_0}{a_B}
(673.7 d_p -436.2 d_n) \,,\\
\bra{p_{1/2}} H_{\rm eff}\ket{s_{1/2}}_{{}^{232}{\rm Th}} &= -2 c_{s_{1/2}}c_{p_{1/2}}\frac{\mu_0}{a_B}
(1363 d_p -1473 d_n) \,.
\end{align}
\end{subequations}

In Sect.~\ref{Constraints}, we will compare these quantities with the matrix element of the contact interaction (\ref{matrix-contact}) and will find the limits on the nucleon EDMs implied by the experimental constraints on $C_{SP}$. However, before proceeding further, let us make a few comments about the details and the accuracy of our calculation of the matrix elements \eqref{317}.

The leading contributions to the atomic EDM from nucleon EDMs arise from $s_{1/2}$ - $p_{1/2}$ electron matrix elements, because these states are significantly enhanced near the nucleus by the absence of the centrifugal barrier and relativistic effects \cite{khriplovich1991parity}. Therefore, we ignore contributions from matrix elements with higher-$l$ wave functions which are very small near the nucleus ($\propto r^l$).

We call attention to the fact that the effective electron-nucleon interactions \eqref{EffectiveOps} fall with the distance (similar to the effect of the ordinary nuclear polarizability) because they have dipole nature. These interactions give the main contributions to the matrix elements at distances from about nuclear radius to hundreds of nuclear radii. Thus, it is localised in the near-nucleus region with $R\ll a_B/Z^{1/3}$, where $a_B$ is the Bohr radius. It is important to note that in this region the inter-electron interaction and screening are negligible, and we can use approximate electron wave functions (\ref{B2}) for $s_{1/2}$ and $p_{1/2}$ bound states. The virtual excited electronic states are described by Dirac-Coulomb wave functions in the continuum spectrum (\ref{HyperGeo}) while the excited states in the discrete spectrum give negligible contributions. For those radial wave functions, which are singular at the origin for a point-like nucleus, we consider their regular polynomial continuation inside the nucleus such that at the origin they behave similarly to the solutions for the Dirac particle inside of a constantly charged ball. This approximation does not break the accuracy of our results.

The radial integrals in the electronic matrix elements are calculated numerically. To control the accuracy of numerical integration, we checked our numerical methods by calculating the energy shifts (contributions to the Lamb shifts) in heavy one-electron ions due to the nuclear polarizability. The error in this calculation does not exceed 5\% as compared with earlier calculations for Th, U, Pu, Cm and other heavy atoms \cite{Plunien91,Plunien95}. Technically, this energy shift is characterised by electronic matrix elements with electric electron-nucleon interaction while in our work we have similar matrix elements with combined electric and magnetic interaction. This comparison serves as a good check of the accuracy of our numerical methods in the electronic part of the calculations. However, the main error in Eq.~(\ref{317}) stems from the quantities $M_p$ and $M_n$ where we expect the errors of our calculations for the nuclear matrix elements and corresponding energies to be under 50\%. This estimate is based on the comparison of the reduced transition probabilities for M1 spin-flip transitions calculated within the single-particle nuclear shell model employed in our work with the results of more sophisticated nuclear calculations presented in Ref.~\cite{nuc-res} for some heavy nuclei (see comments at the end of Appendix~\ref{NuclearSection}). Thus, we conclude that the error in our calculation of the matrix elements \eqref{317} is within 50\% for all atoms. This level of accuracy is typical for most of calculations of $CP$-violating hadronic effects quoted, e.g., in Eqs.~(\ref{dp-theta}) and (\ref{gandtheta}). Further errors originating from relations between hadronic $CP$-violating parameters will be introduced into the calculation later when we derive relations between $C_{SP}$ and these parameters. Note also that constraints on hadronic $CP$-violating parameters in heavy atoms have usually logarithmic scale where the accuracy of up to a factor of two is acceptable.

%%%%%%%%%%%%%%%%%%%%%%%%%%%%%%%%%%%%%%%%%%%%%%%%%%%%%%%%%%%%%%%%%%%%%%%

\section{Constraints on \texorpdfstring{$CP$}{}-odd hadronic parameters}\label{Constraints}

In this section, we will compare the matrix element (\ref{matrix-d}) with the corresponding matrix element of the contact interaction (\ref{matrix-contact}). This will allow us to determine the dependence of the coupling constant $C_{SP}$ on the nucleon permanent EDMs $d_p$ and $d_n$. Then, employing the experimental constraint (\ref{Csp-constr}) we will determine the limits on the parameters $d_{p}$ and $d_n$ originating from the EDM experiments with paramagnetic molecules. We will also present the constraints on other hadronic parameters, which follow from the obtained ones.

\subsection{Limits on nucleon EDMs}
In Eq.~\eqref{matrix-d}, we computed the contributions to the atomic EDM \eqref{de_contact} due to the nucleon EDMs. It is natural to compare these contributions with that due to the contact interaction $H_{\rm cont}$. Equating the matrix elements of these interactions (\ref{matrix-contact}) and (\ref{matrix-d}) we find the relation between the constant $C_{SP}$ and the parameters $d_p$ and $d_n$:
\begin{equation}\label{relationspherical}
C^{(d)}_{SP}=\frac{20\sqrt2}{G_F}\frac{\Gamma(2\gamma+1)^2}{1+4\gamma}\frac{\pi\mu_0 R_0^2}{AZ\alpha}\left(\frac{a_B}{2ZR_0}\right)^{2\gamma}\left[ d_p (g^s_p-\epsilon g^l_p) M_p +d_n (g^s_n-\epsilon g^l_n) M_n \right]\,.
\end{equation}
We point out that the right-hand side of Eq.\ \eqref{relationspherical} is independent of the wave functions normalization constants $c_{s_{1/2}}$ and $c_{p_{1/2}}$.

We stress that the contribution to $C_{SP}$ \eqref{relationspherical} originates from discrete nuclear intermediate excited states while the authors of Ref.~\cite{FPRS} found analogous contributions due to virtual nuclear excitations to continuum spectrum,
\begin{equation}\label{Pospelov's contribution}
    C^{(c)}_{SP}=\frac{8\sqrt{2}m_e\alpha^2{\rm ln}(A)}{G_Fp_Fm_N}\left(\frac{Z}{A}\frac{\mu_p}{\mu_0}\frac{d_p}{e}+\frac{N}{A}\frac{\mu_n}{\mu_0}\frac{d_n}{e}\right)\,.
\end{equation}
Here $p_F\approx250$ MeV is the Fermi momentum of the nucleus, $m_N$ is the nucleon mass and $\mu_{p}$, $\mu_n$ are the nucleon magnetic moments  ($\mu_p/\mu_0=2.793$, $\mu_n/\mu_0=-1.913$). The superscripts $(d)$ and $(c)$ in Eqs.~\eqref{relationspherical} and \eqref{Pospelov's contribution} refer to the contributions of discrete and continuum intermediate nuclear states, respectively.

Combining the two contributions \eqref{relationspherical} and \eqref{Pospelov's contribution}, we find the leading-order dependence of $C_{SP}=C^{(d)}_{SP}+C^{(c)}_{SP}$ on nucleon EDMs,
\begin{equation}
C_{SP}=(\lambda_1d_p+\lambda_2d_n)\times \frac{10^{13}}{e\cdot{\rm cm}}\,,
\label{num-relation}
\end{equation}
where the numerical values of the coefficients $\lambda_1$ and $\lambda_2$ for several atoms of interest are presented in Table\ \ref{Results}. 

The relation \eqref{num-relation} represents the central results of this paper.
It allows us to place limits on the nucleon EDMs. Taking into account the most recent constraints on $|C_{SP}|$ from the ${}^{180}$HfF$^+$ \cite{HfF2017} and ${}^{232}$ThO \cite{ACMEII} EDM experiments (see also Ref.~\cite{fleig2018}), we obtain the limits on $d_p$ and $d_n$. These results are collected in Table \ref{Limits} at the end of this section.

We note that since the quantities $\lambda_{1,2}$ and the corresponding limits on $d_{p,n}$ are directly linked to the effective matrix elements $M_{p,n}$, the errors in these quantities are similar to those incurred in the computation of $M_{p,n}$, i.e., around 50\%.

\subsection{Limits on \texorpdfstring{$CP$}{}-odd pion-nucleon coupling constants} 

The nucleon EDM may appear due to more fundamental interactions. In this subsection, we assume that the dominant contribution to the nucleon EDM is stipulated by the $CP$-odd pion-nucleon interaction.

The $CP$-odd pion-nucleon interactions with coupling constants $\bar g^{(0)}_{\pi NN}$ and $\bar g^{(1)}_{\pi NN}$ may induce atomic EDMs through different mechanisms: (i) they contribute to the nucleon EDMs via one-loop quantum corrections \cite{POSPELOV2005} and (ii) they contribute to the effective nucleon-electron interaction through one- and two-loop contributions to the nucleon polarizability \cite{FPRS}. In this section, we revisit all these contributions and combine them with those from $d_p$ and $d_n$ found in the previous subsection. This will allow us to find the constraints on $\bar g^{(0)}_{\pi NN}$ and $\bar g^{(1)}_{\pi NN}$ originating from the EDM experiments with paramagnetic atoms and molecules. 

We recall that the coupling constants $\bar g^{(0)}_{\pi NN}$ and $\bar g^{(1)}_{\pi NN}$ enter the $CP$-odd pion-nucleon interaction as follows:
\begin{equation}
    {\cal L}_{\pi NN} = \bar g_{\pi NN}^{(0)} \bar N \tau^a N \pi^a
    + \bar g_{\pi NN}^{(1)} \bar N N \pi^0 \,,
\label{L}
\end{equation}
where $N$ is the nucleon doublet, $\pi^a,\pi^0$ are the pion fields, and $\tau^a$ are isospin Pauli matrices. This interaction is responsible for the one-loop quantum correction to the nucleon EDM which were originally estimated in Ref.~\cite{CVVW1979} and revisited in subsequent papers \cite{Hockings,Ottnad,Guo},
\begin{subequations}
\label{dp-g-relation}
\begin{eqnarray}
    d_n &\approx& \frac{e g_A \bar g^{(0)}_{\pi NN}}{8\pi^2 F_\pi}
    \left( 2\ln \frac{m_p}{m_\pi} +\frac\pi2\frac{m_\pi}{m_p}-1\right)    \approx 1.1 \times 10^{-14} \bar g^{(0)}_{\pi NN} \,e\cdot{\rm cm}\,,\\
    d_p&\approx&\frac{e g_A \bar g^{(0)}_{\pi NN}}{8\pi^2 F_\pi}
    \left( 1-2\pi \frac{m_\pi}{m_p}-2 \ln\frac{m_p}{m_\pi} \right)\approx -1.3 \times 10^{-14} \bar g^{(0)}_{\pi NN} \,e\cdot{\rm cm}\,,
\end{eqnarray}
\end{subequations}
where $g_A\approx1.3$ is the axial triplet coupling constant, $F_\pi \approx 93 $ MeV is the pion decay constant, $m_\pi \approx 135 $ MeV is the pion mass and $m_p\approx m_n\approx 938$~MeV is the nucleon mass.
In Eqs.~(\ref{dp-g-relation}) we keep only the terms originating from the meson cloud and omit the (counter)terms corresponding to short-distance effects (see, e.g., \cite{Ottnad,Guo,yamanaka2017}). The latter terms depend on low-energy constants with unknown values which bring a high level of uncertainty in the relations (\ref{dp-g-relation}). An accurate study of these short-distance effects goes beyond the scope of this paper. 

There are one- and two-loop quantum contributions to the nucleon polarizability $\beta_{p,n}=\beta_{p,n}^{(1)}+\beta_{p,n}^{(2)}$, which involve interaction vertices (\ref{L}) \cite{FPRS} \footnote{In Eqs.~(\ref{nucl-pol-a}) and (\ref{nucl-pol-b}), the last terms account for the contributions from $\eta$ meson exchange with the mass $m_\eta$. These mesons have $CP$-odd meson-nucleon interaction with coupling constant $\bar g_{0,\eta}\approx 5\bar g_{0,\pi}$.}:
\begin{subequations}
\label{nucl-pol}
\begin{align}
\beta^{(1)}_p &= -\frac{\alpha}{\pi F_\pi m_\pi^2}
 \left( \bar g^{(1)}_{\pi NN} + \bar g^{(0)}_{\pi NN} + \frac5{\sqrt3}\frac{m_\pi^2}{m_\eta^2} \bar g^{(0)}_{\pi NN} \right)\,,\label{nucl-pol-a}\\
\beta^{(1)}_n &= -\frac{\alpha}{\pi F_\pi m_\pi^2}
 \left( \bar g^{(1)}_{\pi NN} - \bar g^{(0)}_{\pi NN} + \frac5{\sqrt3}\frac{m_\pi^2}{m_\eta^2} \bar g^{(0)}_{\pi NN} \right)\,,\label{nucl-pol-b}\\
\beta_p^{(2)} &= - \frac{\alpha g_A \bar g^{(0)}_{\pi NN}}{4F_\pi m_n m_\pi} \frac{\mu_n}{\mu_0}\,,\\
\beta_n^{(2)} &=  \frac{\alpha g_A \bar g^{(0)}_{\pi NN}}{4F_\pi m_p m_\pi} \frac{\mu_p}{\mu_0}\,.
\end{align}
\end{subequations}
Note that these relations contain only the leading terms which preserve the isospin symmetry.
The nucleon polarizability corrections (\ref{nucl-pol}) amount to the following contributions to the effective coupling constant $C_{SP}$
\begin{equation}
\label{Cbeta}
    C_{SP}^{(\beta)} = -\frac{\sqrt2}{G_F}
    \left(
     \frac Z A \beta_p + \frac N A \beta_n
    \right) \frac{3\alpha m_e}{2\pi} \ln\frac{\Lambda}{m_e}\,,
\end{equation}
where the renormalization scale $\Lambda = m_\rho\approx 775$ MeV for one-loop corrections and $\Lambda=m_\pi\approx135$ MeV for the two-loop ones (see \cite{FPRS} for detail).  

Now we substitute the relation (\ref{dp-g-relation}) into Eq.~(\ref{relationspherical}) or (\ref{num-relation}) and add also the contribution (\ref{Cbeta}). As a result, we find the leading-order relation between the constant $C_{SP}$ and $CP$-violating pion-nucleon couplings
\begin{equation}
    C_{SP}=\lambda_3\bar{g}^{(0)}_{\pi NN}+\lambda_4\bar{g}^{(1)}_{\pi NN}\,,
    \label{2}
\end{equation}
where the numerical values of the coefficients $\lambda_3$ and $\lambda_4$ are collected in Table~\ref{Results} for different atoms. The corresponding limits on these couplings originating from the experimental constraints are given in Table~\ref{Limits} below. We stress that these limits are based on the assumption that the $CP$-odd pion interaction (\ref{L}) gives dominant contribution to the atomic EDM.

Unfortunately, we cannot accurately estimate computational errors of our results for the coefficients $\lambda_3$ and $\lambda_4$ because they are based on the relations \eqref{dp-g-relation} and (\ref{nucl-pol}) with unknown errors. However, by comparing with other analogous calculations (e.g., \cite{Guo}) it is natural to expect that the errors in Eqs.~\eqref{dp-g-relation} and (\ref{nucl-pol}) may be of order $30\%-100\%$. Nevertheless, our relation \eqref{2} is still useful for comparison with other results for $CP$-violation in heavy nuclei, in particular, with the atomic EDMs produced by the nuclear Schiff moments,  which do not contain estimates of the theoretical errors in the nuclear calculations and should be understood on the logarithmic scale.

\subsection{Limits on quark chromo-EDM}

In this subsection, we consider the chromo-EDM of up and down quarks denoted by $\tilde d_u$ and $\tilde d_d$, respectively. Assuming that these quantities are the only sources of the nucleon EDM and $CP$-violating internucleon forces, the authors of Refs.~\cite{PospelovRitz99,FDK,PospelovRitz2001,POSPELOV2005,Abusaif} established the following relations:
\begin{subequations}\label{conversion}
\begin{align}
d_p       &=-(2.1\pm 1.1)e(\tilde{d}_u+0.125\tilde{d}_d)\,,\\
d_n       &=(1.1\pm 0.6) e(\tilde{d}_d+0.5\tilde{d}_u)\,,\\
g\bar{g}^{(0)}_{\pi NN}&=(0.8\pm 0.4)\times 10^{15}(\tilde{d}_u+\tilde{d}_d){\rm cm}^{-1}\,,\label{49c}\\
g\bar{g}^{(1)}_{\pi NN}&=(4.0\pm 2.0)\times 10^{15}(\tilde{d}_u-\tilde{d}_d){\rm cm}^{-1}\,.\label{49d}
\end{align}
\end{subequations}
These equations should be substituted into the sum of Eqs.~(\ref{relationspherical}) and (\ref{Cbeta}), giving the leading-order dependence of $C_{SP}$ on $\tilde d_{u}$ and $\tilde d_p$ as
\begin{equation}
    C_{SP}=(\lambda_5\tilde{d}_u+\lambda_6\tilde{d}_d)\times 10^{15}{\rm cm}^{-1}\,.\label{3}
\end{equation}
The numerical values of the coefficients $\lambda_5$ and $\lambda_6$ are given in Table~\ref{Results}. The corresponding constraints on the values of quark chromo-EDMs are presented in Table~\ref{Limits} below. These constraints are based on the assumption that the quark chromo-EDMs are the dominant sources of nucleon EDM and the $CP$-odd pion-nucleon interaction (\ref{L}). 

In order to estimate the uncertainty in the coefficients $\lambda_{5,6}$ and the corresponding limits on $\tilde{d}_{u,d}$, we note that the relations \eqref{conversion} all have 50\% uncertainty (the error bars of the relations \eqref{49c} and \eqref{49d} were estimated in Ref.\ \cite{yamanaka2017}). Combining these with the 50\% error bars in $\lambda_{1,2}$ and assuming that the relations (\ref{dp-g-relation}) and \eqref{nucl-pol} each have 100\% uncertainty, we find that the values of $\lambda_{5}$ and $\tilde{d}_{u}$ are accurate up to a factor of 3 wheres the values of $\lambda_{6}$ and $\tilde{d}_{d}$ are accurate up to a factor of 2.

\subsection{Limit on QCD vacuum angle \texorpdfstring{$\bar\theta$}{}}

\begin{table}[tb]
\begin{center}
\begin{tabular}{|c|c|c|c|c|c|c|c|c|}
\hline
\multicolumn{2}{|c|}{} & $\lambda_1$ & $\lambda_2$ & $\lambda_3$ & $\lambda_4$ & $\lambda_5$ & $\lambda_6$ & $\lambda_7$ \\
 \hline
\multirow{3}{*}{\rotatebox[origin=c]{90}{Spherical}}         & ${}_{56}^{138}{\rm Ba}$ & 6.4 & -7.2 & -2.3 & 2.2 & 0.4 & -0.8 & 5.1 \\[0.25ex] \cline{2-9}
                                                             & ${}_{82}^{206}{\rm Pb}$ & 5.5 & -6.2 & -2.1 & 2.2 & 0.5 & -0.8 & 4.6 \\[0.25ex] \cline{2-9}
                                                             & ${}_{82}^{208}{\rm Pb}$ & 5.5 & -5.3 & -2.0 & 2.2 & 0.5 & -0.8 & 4.4 \\[0.25ex] \hline
\multirow{7}{*}[-2.5ex]{\rotatebox[origin=c]{90}{Deformed}}  & ${}_{70}^{172}{\rm Yb}$ & 9.4 & -8.4 & -2.8 & 2.2 & 0.4 & -0.8 & 6.0 \\[0.25ex] \cline{2-9}
                                                             & ${}_{70}^{174}{\rm Yb}$ & 9.3 & -9.7 & -3.0 & 2.2 & 0.4 & -0.8 & 6.3 \\[0.25ex] \cline{2-9}
                                                             & ${}_{70}^{176}{\rm Yb}$ & 9.2 & -8.6 & -2.8 & 2.2 & 0.4 & -0.8 & 6.1 \\[0.25ex] \cline{2-9}
                                                             & ${}_{72}^{178}{\rm Hf}$ & 12  & -8.2 & -3.1 & 2.2 & 0.3 & -0.8 & 6.5 \\[0.25ex] \cline{2-9}
                                                             & ${}_{72}^{180}{\rm Hf}$ & 12  & -9.0 & -3.2 & 2.2 & 0.3 & -0.8 & 6.7 \\[0.25ex] \cline{2-9}
                                                             & ${}_{88}^{226}{\rm Ra}$ & 5.7 & -5.7 & -2.1 & 2.2 & 0.5 & -0.8 & 4.6 \\[0.25ex] \cline{2-9}
                                                             & ${}_{90}^{232}{\rm Th}$ & 6.4 & -6.9 & -2.3 & 2.2 & 0.4 & -0.8 & 5.1 \\[0.25ex] \hline
\end{tabular}
\caption{\label{Results}
The results of numerical computations of the coefficients $\lambda_1,\ldots ,\lambda_{7}$ in Eqs.~\eqref{num-relation}, (\ref2), (\ref3) and (\ref4).}
\end{center}
\end{table}
\begin{table}[tb]
\begin{center}
\begin{tabular}{|c|c|c|}
\hline
                 			  & $^{232}{\rm ThO}$                   & $^{180}{\rm HfF}^+$ \\ \hline
$|C_{SP}|$       			  & $7.3\times 10^{-10}$\ \cite{ACMEII} & $1.8\times 10^{-8}$\ \cite{HfF2017,fleig2018} \\ \hline
$|d_p|$                       & $1.1\times 10^{-23}e\cdot{\rm cm}$  & $1.5\times 10^{-22}e\cdot{\rm cm}$ \\ \hline
$|d_n|$                       & $1.0\times 10^{-23}e\cdot{\rm cm}$  & $2.0\times 10^{-22}e\cdot{\rm cm}$ \\ \hline
$|\bar{g}^{(0)}_{\pi NN}|$    & $3.1\times 10^{-10}$                & $5.6\times 10^{-9}$                 \\ \hline
$|\bar{g}^{(1)}_{\pi NN}|$    & $3.3\times 10^{-10}$                & $8.2\times 10^{-9}$                 \\ \hline
$|\tilde{d}_d|$  			  & $9.3\times 10^{-25}{\rm cm}$        & $2.2\times 10^{-23}{\rm cm}$        \\ \hline
$|\tilde{d}_u|$  			  & $1.7\times 10^{-24}{\rm cm}$        & $5.8\times 10^{-23}{\rm cm}$        \\ \hline
$|\bar{\theta}|$ 			  & $1.4\times 10^{-8}$                 & $2.7\times 10^{-7}$                 \\ \hline
\end{tabular}
\end{center}
\caption{\label{Limits} Limits on absolute values of $CP$-violating hadronic parameters arising from the relations \eqref{num-relation}, (\ref2), (\ref3) and (\ref4) upon implementing the constraints on the constant $C_{SP}$ from the $^{180}{\rm HfF}^+$ \cite{HfF2017,fleig2018} and $^{232}{\rm ThO}$ \cite{ACMEII} experiments.}
\end{table}

The QCD vacuum angle $\bar\theta$ is the fundamental $CP$-odd parameter which can induce the nucleon (and atomic) EDM. In this subsection, we find the limit on $\bar\theta$ in the assumption that the nucleon EDM is dominated by contributions from the QCD vacuum angle.

The dependence of the nucleon EDMs on $\bar\theta$ was originally estimated in Ref.~\cite{CVVW1979} and refined in subsequent papers \cite{PospelovRitz99,deVries2015,bsaisou2015,deVries2016}:
\begin{equation}
\label{dp-theta}
d_p =(2.1\pm1.2)\times10^{-16}\bar{\theta}\, e\cdot{\rm cm}\,,\qquad
d_n =-(2.7\pm1.2)\times10^{-16}\bar{\theta}\, e\cdot{\rm cm}\,.
\end{equation}
The pion-nucleon coupling constants $\bar g^{(0)}_{\pi NN}$ and $\bar g^{(1)}_{\pi NN}$ can also be expressed via $\bar\theta$ as \cite{POSPELOV2005,deVries2015,Bsaisou2012} (see also Refs.~\cite{deVries2016,yamanaka2017} for reviews\footnote{We point out that the signs in the relations (103) and (104) between $\bar g^{(0,1)}_{\pi NN}$ and $\bar \theta$ quoted in Ref.~\cite{yamanaka2017} are inconsistent with the definition of ${\cal L}_{\pi NN}$ presented in that review.})
\begin{equation}\label{gandtheta}
\bar{g}^{(0)}_{\pi NN}=-(15.5\pm2.5)\times 10^{-3}\,\bar{\theta}\,,\qquad
\bar{g}^{(1)}_{\pi NN}=(3.4\pm2)\times 10^{-3}\,\bar{\theta}\,.
\end{equation}
Substituting these relations into the sum of Eqs.~(\ref{relationspherical}) and (\ref{Cbeta}) we represent $C_{SP}$ in terms of $\bar\theta$
\begin{equation}
    C_{SP}=\lambda_{7}\times10^{-2}\bar{\theta}\label{4}\,,
\end{equation}
where the value of the constant $\lambda_{7}$ is given in Table~\ref{Results} for different atoms. In particular, with the use of the corresponding value for $^{232}$Th, the experimental constraint on $C_{SP}$ (\ref{Csp-constr}) implies 
\begin{equation}
\label{bar-theta-constr}
    |\bar\theta| < 1.4\times 10^{-8}\,.
\end{equation}
This constraint is close to the result advocated in the recent paper \cite{FPRS}. 

Using the uncertainties in Eqs.~(\ref{dp-theta}) and (\ref{gandtheta}) and assuming that the relations (\ref{dp-g-relation}) and  \eqref{nucl-pol} each have 100\% uncertainty, we find that the error bars in the relation \eqref{4} and the corresponding limit \eqref{bar-theta-constr} are about $70\%$.

\section{Summary and discussion}\label{Conclusion}

In this paper, we demonstrated that the experiments measuring the electron electric dipole moment with paramagnetic atoms and molecules are also sensitive to nucleon EDMs. Dominant contributions to the atomic EDM in such atoms arise from the combined electric and magnetic electron-nucleus interaction. Taking into account nuclear structure effects, we derived the leading-order relations (\ref{num-relation}) between the electron-nucleus contact interaction constant $C_{SP}$ and nucleon permanent EDMs $d_p$ and $d_n$. As a result, the constraint (\ref{Csp-constr}) on the parameter $C_{SP}$ allows us to find limits on nucleon EDMs arising from the experiments with paramagnetic molecules: 
\begin{equation}
    |d_p|<1.1\times10^{-23}e\cdot{\rm cm}\,,\qquad
    |d_n|<1.0\times10^{-23}e\cdot{\rm cm}\,.
\label{Concl-limits}
\end{equation}

It is instructive to compare these limits with the currently accepted ones. In particular, our limit on the neutron EDM obtained from the results of experiments with paramagnetic molecules is almost three order of magnitude weaker than the recent experimental measurements of EDM of neutron $|d_n|<1.8\times 10^{-26} e\cdot{\rm cm}$ \cite{EDMneutron}. However, for the proton EDM our limit (\ref{Concl-limits}) is just about 20 times weaker than the recent constraint on this parameter $|d_p|<5\times 10^{-25}e\cdot{\rm cm}$ \cite{FD2020} which was based on the measurements on EDM of $^{199}$Hg atom \cite{Graner2016}.\footnote{Note that this constraint is approximately two times weaker than that in Ref.~\cite{Sahoo} because the authors of Ref.~\cite{FD2020} revisited earlier calculations of nuclear Shiff moments.} Remarkably, the constraint (\ref{Concl-limits}) on the proton EDM is nearly 30 times more stringent than that found in recent $^{129}$Xe EDM experiments \cite{Xe1,Xe2}. This sensitivity of eEDM experiments to hadronic $CP$-violating parameters is actually very impressive. We expect that further improvement of accuracy in the experiments with paramagnetic molecules would push these limits.

The nucleon EDM may be expressed via more fundamental $CP$-violating parameters such as $CP$-odd pion-nucleon coupling constants $\bar g^{(0,1)}_{\pi NN}$, quark chromo-EDMs $\tilde d_{u,d}$ and the QCD vacuum angle $\bar\theta$. This allows us to find the leading-order dependence of the contact interaction coupling constant $C_{SP}$ on these parameters. The results are represented by Eqs.~(\ref{2}), (\ref{3}) and (\ref{4}) with coefficients $\lambda_3,\ldots, \lambda_7$ given in Table \ref{Results}. The corresponding limits arising on these parameters from the experiments \cite{HfF2017} and \cite{ACMEII} are presented in Table~\ref{Limits}. In particular, the QCD vacuum angle is limited as $|\bar\theta|<1.4\times10^{-8}$ which is approximately two orders of magnitude weaker than the currently accepted constraint from neutron and Hg atom EDM experiments $|\bar\theta|<10^{-10}$ \cite{PDG}, but is nearly three times more stringent than the corresponding bounds from $^{129}$Xe EDM experiments \cite{Xe1,Xe2}.
Note that the limits presented above do not include theoretical errors which often are unknown (see, e.g., calculation of the $^{199}$Hg nuclear Schiff moment in Ref.~\cite{Engel} where different interaction models give different sign of the result).
We provided error estimates for all obtained results where appropriate.
In particular, for the limits on the nucleon EDMs $d_{p,n}$ the errors do not exceed 50\%,
the relation for the QCD vacuum angle $\bar \theta$ has about 70\% uncertainty while the limit on the up quark chromo-EDM $\tilde d_u$ is accurate up to a factor of 3 and the limit on the down quark chromo-EDM $\tilde d_d$ is accurate up to a factor of 2.

We stress that obtained in this paper contributions to $C_{SP}$ from nucleon EDMs are independent from and additional to those found in Ref.~\cite{FPRS}, although they originate from the same combined electric and magnetic electron-nucleon interaction. Indeed, the authors of the paper \cite{FPRS} took into account virtual nuclear transitions from bound states to {\it continuum} for all nucleons in the nucleus, so it increases with the nucleon number $A$.  In our paper, in contrast, we consider virtual nuclear transitions to excited {\it bound} nuclear states for several external shell nucleons which can flip their spins. This contribution is enhanced by the small energy denominators and large matrix elements of the spin operator between the spin-orbit doublet components.  As we demonstrate in this paper, accounting these M1 nuclear spin-flip transitions approximately doubles the results presented in Ref.~\cite{FPRS}. 

To conclude, we expect that the obtained in this paper results may place more stringent limits on $CP$-violating hadronic parameters once improved constraints on $C_{SP}$ are available from next generations of eEDM experiments \cite{Baron2014,ACMEII,Hudson2011,Loh1220,HfF2017,BaF2018}. Note that the sensitivity in these experiments improved by two orders of magnitude during the last decade.

\section*{Acknowledgements}
This work was supported by the Australian Research Council Grants No. DP150101405 and DP200100150 and the Gutenberg Fellowship. We thank Vladimir Dmitriev, Maxim Pospelov, Adam Ritz, Yevgeny Stadnik and Anna Viatkina for useful discussions.

\appendix
\section{Nuclear energies and matrix elements}\label{NuclearSection}

In this appendix, we estimate the matrix elements and corresponding energies of nuclear M1 spin-flip single-particle transitions. The details of these computations slightly differ for (nearly) spherical and deformed nuclei. Therefore, we consider these two cases separately.

\subsection{Spherical nuclei}

In this section, we focus on the $^{208}$Pb, $^{206}$Pb and $^{138}$Ba nuclei, which are nearly spherical, i.e., they have deformation $\delta<0.1$. For these nuclei, proton and neutron single-particle states may be labeled as $|n,l,j,m\rangle$, where $n$ is the oscillator quantum number, $l$ and $j$ are the orbital and total momentum numbers, $m$ is magnetic quantum number. In this basis, the nuclear spin operator $\bf s$ provides transitions between fine structure doublets. 

\begin{table}
\small
\begin{center}
\begin{tabular}{|c|c|c|c|c|c|c|}
\hline
\multirow{2}{*}{} & \multicolumn{2}{c}{Proton transitions} & 
\multicolumn{2}{|c|}{Neutron transitions} & $R_0$& \multirow{2}{*}{$\delta$} \\ \cline{2-5}
& $|\langle n' |{\bf s}|0'\rangle_p |^2$ & $\Delta E_{n'}$ (MeV) & $|\langle n' |{\bf s}|0\rangle_n' |^2$ & $\Delta E_{n'}$ (MeV) & (fm) & \\ \hline
\multirow{7}{*}{\rotatebox[origin=c]{90}{$^{138}$Ba}} & 18/25 & 2.7 & 170/121 & 5.3 & \multirow{7}{*}{6.20} & \multirow{7}{*}{0.09}\\
                            						  & 2/25    & 4.1 & 200/121 & 5.4 & & \\
                            						  & 28/81   & 4.3 & 30/121  & 5.5 & & \\
                           							  & 56/81   & 4.4 & 136/121 & 5.9 & & \\
                            						  & 16/81   & 4.5 & 56/121  & 6.0 & & \\
                            						  & 8/9     & 4.6 & 60/121  & 6.2 & & \\
                            						  & 8/81    & 5.2 & 8/121   & 6.5 & & \\
\hline
\multirow{9}{*}{\rotatebox[origin=c]{90}{$^{206}$Pb}} & 10/11   & 4.5 & 72/169  & 6.1 & \multirow{9}{*}{7.09} & \multirow{9}{*}{0.03} \\
                                                      & 162/121 & 4.6 & 462/169 & 6.2 & & \\
                                                      & 98/121  & 4.7 & 318/169 & 6.3 & & \\
                            						  & 250/121 & 4.8 & 132/169 & 6.4 & & \\
                            						  & 32/121  & 5.0 & 100/169 & 6.5 & & \\
                            						  & 8/121   & 5.1 & 6/169   & 6.7 & & \\
                           							  &         &     & 2/169   & 6.9 & & \\
                            						  &         &     & 2/3     & 1.4 & & \\
                           							  &         &     & 10/9    & 2.0 & & \\
\hline
\multirow{7}{*}{\rotatebox[origin=c]{90}{$^{208}$Pb}} & 10/11   & 4.5 & 72/169  & 6.1 & \multirow{7}{*}{7.11} & \multirow{7}{*}{0.05} \\
                            						  & 162/121 & 4.6 & 462/169 & 6.2 & & \\
                           							  & 98/121  & 4.7 & 318/169 & 6.3 & & \\
                            						  & 250/121 & 4.8 & 132/169 & 6.4 & & \\
                            						  & 32/121  & 5.0 & 100/169 & 6.5 & & \\
                           							  & 8/121   & 5.1 & 6/169   & 6.7 & & \\
                            						  &         &     & 2/169   & 6.9 & & \\
\hline
\end{tabular}
\caption{Nuclear radii $R_0$, deformation parameters $\delta$, matrix elements $|\langle n' |{\bf s}|0'\rangle_{p,n} |^2$ and the corresponding energies $\Delta E_{n'}$ of M1 spin-flip transitions in some spherical nuclei of interest.}\label{sphericalNuc}
\end{center}
\end{table}

\begin{table}
\begin{center}
\small
\begin{tabular}{|c|c|c|c|c|c|c|}
\hline
\multirow{2}{*}{} & \multicolumn{2}{c}{Proton transitions} & 
\multicolumn{2}{|c|}{Neutron transitions} & $R_0$ & \multirow{2}{*}{$\delta$} \\ \cline{2-5}
& Transition & $\Delta E_{n'}$ (MeV) & Transition & $\Delta E_{n'}$ (MeV) & (fm) &\\ \hline
\multirow{5}{*}{\rotatebox[origin=c]{90}{$^{172}$Yb}} & $\ket{523\frac72} \to  \ket{523\frac52} $ & 4.5 & $ \ket{651\frac{3}2} \to  \ket{651\frac12}$ & 3.9 & \multirow{5}{*}{6.67} & \multirow{5}{*}{0.31} \\
                                                      & $\ket{532\frac52} \to  \ket{532\frac32} $ & 4.0 & $ \ket{642\frac52} \to  \ket{642\frac32}$                  & 4.5 & &\\
                                                      & $\ket{541\frac32} \to  \ket{541\frac12} $ & 4.5 & $ \ket{633\frac72} \to  \ket{633\frac52}$                  & 5.0 & &\\
                                                      & $\ket{404\frac92} \to  \ket{404\frac72} $ & 4.1 & $ \ket{505\frac{11}2} \to  \ket{505\frac92}$               & 5.1 & &\\
                                                      &                                        &     & $ \ket{514\frac{9}2} \to  \ket{514\frac{7}2}$              & 4.6 & &\\
\hline
\multirow{6}{*}{\rotatebox[origin=c]{90}{$^{174}$Yb}} & $\ket{523\frac72} \to  \ket{523\frac52} $ & 4.5 & $ \ket{651\frac{3}2} \to  \ket{651\frac12}$ & 3.9 & \multirow{6}{*}{6.70} & \multirow{6}{*}{0.31} \\
                                                      & $\ket{532\frac52} \to  \ket{532\frac32} $ & 4.0 & $ \ket{642\frac52} \to  \ket{642\frac32}$                  & 4.5 & &\\
                                                      & $\ket{541\frac32} \to  \ket{541\frac12} $ & 4.5 & $ \ket{633\frac72} \to  \ket{633\frac52}$                  & 5.0 & &\\
                                                      & $\ket{404\frac92} \to  \ket{404\frac72} $ & 4.1 & $ \ket{505\frac{11}2} \to  \ket{505\frac92}$               & 5.1 & &\\
                                                      &                                        &     & $ \ket{514\frac{9}2} \to  \ket{514\frac{7}2}$              & 4.6 & &\\
                                                      &                                        &     & $ \ket{512\frac{5}2} \to  \ket{512\frac{3}2}$              & 2.4 & &\\
\hline
\multirow{5}{*}{\rotatebox[origin=c]{90}{$^{176}$Yb}} & $\ket{523\frac72} \to  \ket{523\frac52} $ & 4.5 &
    $ \ket{651\frac{3}2} \to  \ket{651\frac12}$ & 4.2 & \multirow{5}{*}{6.72} & \multirow{5}{*}{0.29} \\
                                                      & $\ket{532\frac52} \to  \ket{532\frac32} $ & 4.1 & $ \ket{642\frac52} \to  \ket{642\frac32}$                  & 4.5 & &\\
													  & $\ket{541\frac32} \to  \ket{541\frac12} $ & 4.5 & $ \ket{633\frac72} \to  \ket{633\frac52}$                  & 5.0 & &\\
													  & $\ket{404\frac92} \to  \ket{404\frac72} $ & 4.0 & $ \ket{505\frac{11}2} \to  \ket{505\frac92}$               & 5.2 & &\\
													  &                                        &     & $ \ket{512\frac{5}2} \to  \ket{512\frac{3}2}$              & 2.4 & &\\
\hline
\multirow{5}{*}{\rotatebox[origin=c]{90}{$^{178}$Hf}} & $\ket{523\frac72} \to  \ket{523\frac52} $ & 4.4 & $ \ket{505\frac{11}2} \to  \ket{505\frac92}$ & 4.2 & \multirow{5}{*}{6.75} & \multirow{5}{*}{0.26} \\
                                                      & $\ket{532\frac52} \to  \ket{532\frac32} $ & 4.1 & $ \ket{512\frac52} \to  \ket{512\frac32}$                  & 2.4 & &\\
													  & $\ket{541\frac32} \to  \ket{541\frac12} $ & 4.1 & $ \ket{633\frac72} \to  \ket{633\frac52}$                  & 5.0 & &\\
													  & $\ket{402\frac52} \to  \ket{402\frac32} $ & 1.9 & $ \ket{642\frac52} \to  \ket{642\frac32}$                  & 4.6 & &\\
												      & $\ket{411\frac32}\to\ket{411\frac12}$     & 1.4 & $ \ket{631\frac{3}2} \to  \ket{631\frac12}$                & 7.8 & &\\
\hline
\multirow{6}{*}{\rotatebox[origin=c]{90}{$^{180}$Hf}} & $\ket{523\frac72} \to  \ket{523\frac52} $ & 4.4 & $ \ket{505\frac{11}2} \to  \ket{505\frac92}$ & 4.2 & \multirow{6}{*}{6.80} & \multirow{6}{*}{0.25} \\
													  & $\ket{532\frac52} \to  \ket{532\frac32} $ & 4.1 & $ \ket{512\frac52} \to  \ket{512\frac32}$                  & 2.4 & &\\
													  & $\ket{541\frac32} \to  \ket{541\frac12} $ & 4.1 & $ \ket{624\frac92} \to  \ket{624\frac72}$                  & 5.3 & &\\
													  & $\ket{402\frac52} \to  \ket{402\frac32} $ & 1.9 & $ \ket{633\frac72} \to  \ket{633\frac52}$ 				  & 5.0 & &\\
													  & $\ket{411\frac32}\to\ket{411\frac12}$     & 1.4 & $ \ket{642\frac{5}2} \to  \ket{642\frac32}$                & 4.6 & &\\
 													  &                                        &     & $ \ket{631\frac{3}2} \to  \ket{631\frac{1}2}$              & 7.8 & &\\
\hline
\multirow{4}{*}{\rotatebox[origin=c]{90}{$^{226}$Ra}} & $\ket{523\frac72} \to  \ket{523\frac52} $ & 4.3 & $ \ket{624\frac{9}2} \to  \ket{624\frac72}$ & 5.0 & \multirow{4}{*}{7.31} & \multirow{4}{*}{0.20} \\
													  & $\ket{514\frac92} \to  \ket{514\frac72} $ & 4.4 & $ \ket{615\frac{11}2} \to  \ket{615\frac92}$               & 5.0 & &\\
													  & $\ket{505\frac{11}2} \to\ket{505\frac92}$ & 4.4 & $ \ket{606\frac{13}2} \to  \ket{606\frac{11}2}$            & 5.6 & &\\
 													  &                                        &     & $ \ket{761\frac{3}2} \to  \ket{761\frac12}$ 				  & 4.3 & &\\
\hline
\multirow{6}{*}{\rotatebox[origin=c]{90}{$^{232}$Th}} & $\ket{651\frac32} \to  \ket{651\frac12} $  & 4.5 & $ \ket{752\frac52} \to  \ket{752\frac32}$ & 4.1 & \multirow{6}{*}{7.37} & \multirow{6}{*}{0.25} \\
													  & $\ket{505\frac{11}2} \to  \ket{505\frac92} $ & 4.2 & $ \ket{761\frac32} \to  \ket{761\frac12}$                  & 4.0 & &\\
													  & $\ket{514\frac92} \to  \ket{514\frac72} $    & 4.0 & $ \ket{631\frac32} \to  \ket{631\frac12}$                  & 1.0 & &\\
													  & $\ket{523\frac72} \to  \ket{523\frac52} $    & 3.7 & $ \ket{624\frac92} \to  \ket{624\frac72}$                  & 5.0 & &\\
													  &                                           &     & $ \ket{615\frac{11}2} \to  \ket{615\frac92}$               & 4.8 & &\\
													  &  										  &  	& $ \ket{606\frac{13}2} \to  \ket{606\frac{11}2}$            & 5.4 & &\\
\hline
\end{tabular}
\end{center}\caption{Nuclear radii $R_0$, deformation parameters $\delta$ and energies $\Delta E_{n'}$ of M1 spin-flip transitions in some deformed nuclei of interest.}\label{deformNuc}
\end{table}

In the $^{208}$Pb nucleus, the non-vanishing matrix elements of the spin operator are
$\langle 5h\frac92 |{\bf s} |5h\frac{11}{2}\rangle$ for protons and $\langle 6i\frac{11}{2} |{\bf s} |6i\frac{13}{2}\rangle$ for neutrons. The isotope $^{206}$Pb has additional contributions from the $\langle 5p\frac12 |{\bf s}|5p\frac32\rangle$ neutron matrix elements. For $^{138}$Ba, non-vanishing proton contributions arise from the matrix elements $\langle 4d\frac32|{\bf s}|4d\frac52 \rangle$ and $\langle 4g\frac92|{\bf s}|4g\frac72 \rangle$ whereas neutron contributions come from $\langle 5h\frac92 | {\bf s}| 5h\frac{11}2\rangle$. All these matrix elements may be calculated using the properties of spherical spinors (see, e.g., Ref.~\cite{Recurrence}). The energies of all these transitions may be estimated with the use of Fig.~5 in Ref.~\cite{BM}. When the energies are (nearly) degenerate, we give the sum of matrix elements corresponding to the same energy.
In Table~\ref{sphericalNuc} below, we collect the values of such matrix elements with the corresponding energies for $^{208}$Pb, $^{206}$Pb and $^{138}$Ba. The value of the nuclear radius $R_0$ is calculated according to the empirical formula:
\begin{equation}
R_0 = 1.2 A^{1/3}\, {\rm fm}\,.
\end{equation}
For reference, the values of the deformation parameter $\delta$ are also presented.

\subsection{Deformed nuclei}

For deformed heavy nuclei with $\delta>0.1$, it is convenient to use the Nilsson basis \cite{Nilsson,BM}, wherein proton and neutron single-particle states are labeled with $|n_3,n_\perp,\Lambda,\Omega\rangle$, where $n_3$ and $n_\perp$ are the oscillator quantum numbers, $\Lambda$ and $\Omega$ are the projections of angular and total momenta on the deformation axis. Note that $\Omega = \Lambda + \Sigma$ where $\Sigma$ is the projection of the nucleon's spin on the deformation axis. The dependence of the energy levels on the deformation parameter $\delta$ in this model may be inferred from Fig.~5 in Ref.~\cite{BM}. From such dependence, one may estimate the energies of the spin-flip transitions. Note that in the basis $|n_3,n_\perp,\Lambda,\Omega\rangle$, each M1 spin-flip matrix element is $\bra{m'} s_+ \ket{0'} =1$, and the corresponding energy level is doubly degenerate since each quantum number $\Sigma$ corresponds to $\pm \Lambda$. 

The single-nucleon spin-flip M1 transition energies $\Delta E_{n'}$, the deformation parameters $\delta$ and the nuclear radii $R_0$ for several nuclei of interest are presented in Table \ref{deformNuc}.

In conclusions of this subsection we discuss the accuracy of our estimates of matrix elements and corresponding energies presented in Tables~\ref{sphericalNuc} and \ref{deformNuc}. For this purpose, it is convenient to consider the reduced transition probability of M1 spin-flip transition,
$B_0\equiv\sum_{n'} B(M1, 0'\to n')$.
Using the data from Table~\ref{deformNuc}, we find this quantity for $^{232}$Th, $B_0=14.8\mu_0$,
and for $^{172}$Yb, $B_0 = 14.6\mu_0$. 
These values may be compared with the corresponding quantities presented in Ref.~\cite{nuc-res} obtained on the basis of sophisticated Hartree-Fock plus RPA nuclear calculation: $B_0 = 14.9\mu_0$ for $^{232}$Th and $B_0 = 12.8\mu_0$
for $^{172}$Yb. As a result we conclude that the simple single-particle nuclear shell model used in this paper allows us to determine the nuclear M1 matrix elements with error about 15\%.
Errors in nuclear energies are within 50\%.
It may be checked that the accuracy of nuclear calculations for other heavy nuclei considered in our paper is within this range. Therefore, we conclude that the errors in determining values of nuclear matrix elements and corresponding energies of spin-flip M1 transitions are under 50\% for all nuclei. This level of accuracy is acceptable for the goals of this work, although a better accuracy may be achieved with the use of more sophisticated nuclear models.

%%%%%%%%%%%%%%%%%%%%%%%%%%%%%%%%%%%%%%%%%%%%%%%%%%%%%%%%%%%%%%%%%%%%%%%%%%%%%%%%%%%%%%%%%%%%

\section{Evaluation of electronic matrix elements}
\label{AppB}

In this appendix, we provide the details for the numerical calculation of the electronic matrix element \eqref{EMatrix}. For convenience, we use the spherical basis $({\bf e}_+, {\bf e}_-, {\bf e}_0)$. The components of vectors in this basis will be labeled by the $(+,-,0)$ subscripts. Due to spherical symmetry, Eq.\ \eqref{EMatrix} may be rewritten in terms of the `0'-component of the operators $\bf D$ and $\bf M$ introduced in Eqs.~(\ref{EffectiveOps}),
\begin{equation}
M(\Delta E_{n'})
=\frac{\alpha}{c_{s_{1/2}}c_{p_{1/2}}}
\sum_{n}\frac{\bra{p_{1/2}}M_0\ket{n}\bra{n}D_0\ket{s_{1/2}}}{\Delta E_{n}+{\rm sgn}(E_n)\Delta E_{n'}} + (s_{1/2}\leftrightarrow p_{1/2})\,.\label{0}
\end{equation}

In Eq.~\eqref{0}, the sum is taken over all excited electron states $|n\rangle$ with energies $E_n$, including those from the discrete and continuous spectra. As discussed in Sect.~\ref{Contribution to the atomic EDM from nucleon permanent EDMs}, the states from the discrete spectrum give negligible contributions to the electronic matrix elements. Therefore, in what follows, we will consider only intermediate states $|n\rangle$ from the continuum, including both positive and negative energy solutions of the Dirac equation.

For further computation of the matrix element \eqref{0} the electron wave functions need to be specified.

\subsection{The \texorpdfstring{$s_{1/2}$}{} and \texorpdfstring{$p_{1/2}$}{} wave functions}
The valence electron $s_{1/2}$ and $p_{1/2}$ wave functions may be expressed in terms of the spherical spinors $\Omega^\kappa_\mu(\hat{\bf R})$ (see, e.g., Ref.~\cite{Recurrence}) where $\mu$ is the magnetic quantum number and $\kappa=(l-j)(2j+1)$ as
\begin{subequations}
\label{B2}
\begin{align}
\ket{s_{1/2}}&=c_{s_{1/2}}\left(
\begin{array}{c}
f_{s_{1/2}}(R)\Omega^{-1}_{\mu }(\hat{\bf R}) \\
ig_{s_{1/2}}(R)\Omega^{1}_{\mu}(\hat{\bf R})
\end{array}\right)\,,\\
\ket{p_{1/2}}&=c_{p_{1/2}}\left(
\begin{array}{c}
f_{p_{1/2}}(R)\Omega^{1}_{\mu }(\hat{\bf R}) \\
ig_{p_{1/2}}(R)\Omega^{-1}_{\mu}(\hat{\bf R})
\end{array}
\right)\,,
\end{align} 
\end{subequations}
where the radial wave functions $f_{s,p_{1/2}}$ and $g_{s,p_{1/2}}$ are well approximated in the region $R_0<R\ll a_B/Z^{1/3}$ by the Bessel functions of the first kind $J_\nu(x)$ (see, e.g., \cite{khriplovich1991parity}),
\begin{subequations}\label{Bessel}
\begin{align}
f_{s_{1/2}}(R)&=\frac{1}{R}\left[(-1+\gamma)J_{2\gamma}\left(\sqrt{\frac{8ZR}{a_B}}\right)- \frac{1}{2}\sqrt{\frac{8ZR}{a_B}}J_{2\gamma-1}\left(\sqrt{\frac{8ZR}{a_B}}\right)\right]\,,\\
f_{p_{1/2}}(R)&=\frac{1}{R}\left[( 1+\gamma)J_{2\gamma}\left(\sqrt{\frac{8ZR}{a_B}}\right)- \frac{1}{2}\sqrt{\frac{8ZR}{a_B}}J_{2\gamma-1}\left(\sqrt{\frac{8ZR}{a_B}}\right)\right]\,,\\
g_{s_{1/2}}(R)&=g_{p_{1/2}}(R)=\frac{1}{R}Z\alpha J_{2\gamma}\left(\sqrt{\frac{8ZR}{a_B}}\right)\,.
\end{align}
\end{subequations}

Note that the wave functions (\ref{Bessel}) are the zero-energy solutions of the Dirac-Coulomb equations for a point-like nucleus. For an extended nucleus, the corresponding solution is complicated. At the current level of accuracy, it suffices to use Eqs.\ \eqref{Bessel} as an approximation to the wave functions. For the region inside the nucleus, $0\leq R\leq R_0$, the radial wave functions $f_{s,p_{1/2}}$ and $g_{s,p_{1/2}}$ may be continued as follows
\begin{subequations}\label{Bessel-inside}
\begin{align}
f_{s_{1/2}}(R)&=\frac{1}{R_0}\left[(-1+\gamma)J_{2\gamma}\left(\sqrt{\frac{8ZR_0}{a_B}}\right) - \frac{1}{2}\sqrt{\frac{8ZR_0}{a_B}}J_{2\gamma-1}\left(\sqrt{\frac{8ZR_0}{a_B}}\right)\right]\,,\\
f_{p_{1/2}}(R)&=\frac{R}{R_0^2}\left[( 1+\gamma)J_{2\gamma}\left(\sqrt{\frac{8ZR_0}{a_B}}\right) - \frac{1}{2}\sqrt{\frac{8ZR_0}{a_B}}J_{2\gamma-1}\left(\sqrt{\frac{8ZR_0}{a_B}}\right)\right]\,,\\
g_{s_{1/2}}(R)&=\frac{R}{R_0}Z\alpha J_{2\gamma}\left(\sqrt{\frac{8ZR_0}{a_B}}\right)\,,\\
g_{p_{1/2}}(R)&=\frac{1}{R_0}Z\alpha J_{2\gamma}\left(\sqrt{\frac{8ZR_0}{a_B}}\right)\,.
\end{align}
\end{subequations}
Note that these functions are the approximate solutions (containing only leading terms at small distance) of the Dirac equation inside the nucleus with constant density.

Note also that since $8ZR_0/a_B\ll1$, the Bessel functions in Eqs.~(\ref{Bessel-inside}) may be series expanded over their arguments. For computing radial integrals inside the nucleus it is sufficient to keep the leading terms of these expansions as in Eqs.~(\ref{psi-ass}).

\subsection{Excited electronic states of the continuous spectrum}
\label{AppB2}
The excited electronic states $|n\rangle$ in the continuous spectrum may be labeled by the quantum number $\kappa=(l-j)(2j+1)$ and the energy $E$, $|n\rangle \equiv|E\kappa\rangle$. In spherical coordinates, these functions read (see, e.g., Refs.\ \cite{Landau4,greiner2000relativistic}): 
\begin{equation}
\label{B6}
\ket{n}\equiv\ket{E\kappa}=\left(\begin{matrix}
f_{\kappa}^E(R)\Omega^{\kappa}_{\mu}(\hat{\bf R})\\
ig_{\kappa}^E(R)\Omega^{-\kappa}_{\mu}(\hat{\bf R})
\end{matrix}\right)\,,
\end{equation}
with
\begin{subequations}\label{HyperGeo}
\begin{align}
f_{\kappa}^E(R) =&\frac{(2pR)^\gamma e^{\pi y/2}\left|\Gamma(\gamma+iy)\right|\sqrt{\left|E+ m_e\right|}}{R\sqrt{\pi p}\Gamma(2\gamma+1)}\nonumber\\
&\times
{\rm Re}[e^{-ipR+i\eta}{}_1F_1(\gamma+1+iy,2\gamma+1,2ipR)]\,,\label{fCont}\\
g_{\kappa}^E(R) =&-{\rm sgn}(E)\frac{(2pR)^\gamma e^{\pi y/2}\left|\Gamma(\gamma+iy)\right|\sqrt{\left|E- m_e\right|}}{R\sqrt{\pi p}\Gamma(2\gamma+1)}\nonumber\\
&\times{\rm Im}[e^{-ipR+i\eta}{}_1F_1(\gamma+1+iy,2\gamma+1,2ipR)]\,.\label{gCont}
\end{align}
\end{subequations}
Here $p=\sqrt{E^2-m_e^2}$ is the electron's momentum, $y=Z\alpha E/p$, $e^{i\eta}=\sqrt{-\frac{\kappa-iym_e/E}{\gamma+iy}}$ and $_1F_1(a,b,z)$ is the confluent hypergeometric function of the first kind. Note that the wave functions (\ref{B6}) are normalized as $\braket{E'\kappa|E\kappa}=\delta(E'-E)$.

The functions (\ref{HyperGeo}) solve for the Dirac equation with a point-like nucleus. Therefore, we will only use them for outside of the nucleus, $R>R_0$. For the inside of the nucleus, $0\leq R\leq R_0$, we will consider the following continuation of these functions
\begin{equation}\label{Inside}
f^E_\kappa(R)=b_1 R^l\,,\qquad g^E_\kappa(R)=b_2 R^{\tilde{l}}\,,
\end{equation}
where $l=|\kappa+1/2|-1/2$ is the orbital angular momentum corresponding to $\kappa$, $\tilde{l}=|-\kappa+1/2|-1/2$ is the orbital angular momentum corresponding to $-\kappa$. The values of the coefficients $b_{1}$ and $b_2$ are determined by matching Eqs.~(\ref{HyperGeo}) and (\ref{Inside}) on the boundary of the nucleus. The wave functions \eqref{Inside} are, to the leading order, solutions to the Dirac equation inside a nucleus of a constant density.

We stress that the extension of the electronic wave functions to the inside region of the nucleus (\ref{Inside}) is an approximation which is acceptable at our level of accuracy. We checked the validity of this approximation by computing the Lamb shift in heavy atoms due to nuclear polarizability. Within this approximation, we have 95\% agreement with the exact results presented in Refs.~\cite{Plunien91,Pachucki93,Plunien95}.

\subsection{Results of calculation of electronic matrix element}
Substituting the wave functions (\ref{B2}) and (\ref{B6}) into Eq.~(\ref0) and performing the integration over angular variables, we obtain
\begin{equation}
M(\Delta E_{n'})=-\frac{2\alpha }{9}\int\limits_{m_e}^{\infty}\frac{T(E)dE}{E-E_{s_{1/2}}+\Delta E_{n'}}-\frac{2\alpha }{9}\int\limits_{-\infty }^{-m_e}\frac{T(E)dE}{E-E_{s_{1/2}}-\Delta E_{n'}}\,,
\label{EnergyInt}
\end{equation}
where
\begin{equation}
 T(E)=R_s^{(1)}(E)R_p^{(1)}(E)-R_s^{(-2)}(E)R_p^{(-2)}(E)-S_s^{(-1)}(E)S_p^{(-1)}(E)+S_s^{(2)}(E)S_p^{(2)}(E)\,, \label{T(E)} \\ 
\end{equation}
and the radial integrals $R_{s,p}^{(\kappa)}(E)$ and $S_{s,p}^{(\kappa)}(E)$ are defined by
\begin{subequations}
\label{RadialIntegrals}
\begin{align}
 R_s^{(\kappa)}(E)&\equiv \int_0^\infty{\left(f_{s_{1/2}}f_{\kappa}^E+ g_{s_{1/2}}g_{\kappa}^E \right)f( R){{R}^{2}}dR} \,,\\ 
 R_p^{(\kappa)}(E)&\equiv \int_0^\infty{\left(f_{p_{1/2}}g_{\kappa}^E+ g_{p_{1/2}}f_{\kappa}^E  \right)f( R ){{R}^{2}}dR} \,,\\ 
 S_s^{(\kappa)}(E)&\equiv \int_0^\infty{\left(f_{s_{1/2}}g_{\kappa}^E+ g_{s_{1/2}}f_{\kappa}^E \right)f( R ){{R}^{2}}dR} \,,\\ 
 S_p^{(\kappa)}(E)&\equiv \int_0^\infty{\left(f_{p_{1/2}}f_{\kappa}^E+ g_{p_{1/2}}g_{\kappa}^E \right)f( R ){{R}^{2}}dR} \,.
\end{align}
\end{subequations}
Here, the radial function $f(R)$ takes into account the radial dependence of the operators (\ref{EffectiveOps}),
\begin{equation}
f(R)=\theta(R-R_0)\frac{1}{R^2}+\theta(R_0-R)\frac{R}{R_0^3}\,.    
\end{equation}
Note that Eq.~(\ref{T(E)}) involves only the terms with $\kappa=\pm 1,\pm 2$ which are allowed by the selection rules for transitions from $s_{1/2}$ and $p_{1/2}$ bound electron states.

\begin{table}
\renewcommand{\arraystretch}{1.2}
\begin{center}
\begin{tabular}{|c|c|c|c|c|c|c|c|c|c|c|}
\hline
\multirow{2}{*}{} & \multicolumn{3}{|c|}{Spherical} & \multicolumn{7}{|c|}{Deformed}\\
\cline{2-11}
                  & ${}^{138}{\rm Ba}$ & ${}^{206}{\rm Pb}$ & ${}^{208}{\rm Pb}$ & ${}^{172}{\rm Yb}$ & ${}^{174}{\rm Yb}$ & ${}^{176}{\rm Yb}$ & ${}^{178}{\rm Hf}$ & ${}^{180}{\rm Hf}$ & ${}^{226}{\rm Ra}$ & ${}^{232}{\rm Th}$ \\
\hline
$\frac{M_p}{a_B}$             & 11.1 & 94.1 & 94.0  & 69.3 & 69.3 & 69.2 & 121  & 121  & 156  & 244 \\
\hline
$\frac{M_n}{a_B}$             & 16.5 & 143  & 95.7  & 83.4 & 106  & 88.7 & 96.8 & 114  & 196  & 385 \\
\hline
\end{tabular}
\caption{\label{Electronic}
Numerical values for the electronic matrix elements $M_p$ and $M_n$ for several atoms of interest.}
\end{center}
\end{table}

With the radial wave functions (\ref{Bessel}), (\ref{Bessel-inside}), (\ref{HyperGeo}) and (\ref{Inside}), the radial integrals (\ref{RadialIntegrals}) may be computed numerically for any specific electron energy $E$ and nuclear energy $\Delta E_{n'}$. For all values of $\Delta E_{n'}$ presented in Appendix~\ref{NuclearSection}, numerical analysis showed that for $|E|>500m_e$, $T(E)/(E_{s_{1/2}}-E\pm\Delta E_{n'})$ is effectively zero, so the energy integrals in Eqs.~\eqref{EnergyInt} may be cut off at $|E|=500m_e$.  We also point out that the dominant contributions to the energy integrals \eqref{EnergyInt} come from the region where $E\sim 50m_e$, which is larger than the values of $\Delta E_{n'}$ considered in Appendix~\ref{NuclearSection}. As a result, the function $M(\Delta E_{n'})$ has weak energy dependence.

The energy integrals in Eqs.~\eqref{EnergyInt} are computed numerically, giving $M(\Delta E_{n'})$ for all values of $\Delta E_{n'}$ presented in Appendix~\ref{NuclearSection}. The resulting numerical values of the electronic factors $M_p$ and $M_n$ introduced in Eqs.~(\ref{MpMnDef}) are presented in Table.~\ref{Electronic}.

\providecommand{\href}[2]{#2}\begingroup\raggedright\endgroup

\end{document}